\documentclass{llncs}

\makeatletter
\newif\if@restonecol
\makeatother

\usepackage[linesnumbered,ruled,vlined]{algorithm2e}%[ruled,vlined]{
\usepackage{algpseudocode}
\usepackage{amsmath}
\usepackage{amssymb}
\usepackage{graphicx}
\usepackage{CJK}
\usepackage{algorithmicx}
\usepackage{algpseudocode}
\usepackage{amsmath}
\usepackage{subfigure}
\usepackage{float}
\usepackage{url}
\usepackage{array}
\usepackage{booktabs}
\usepackage{threeparttable}
\usepackage{multicol}
\usepackage{multirow}
\usepackage{threeparttable}
\usepackage{enumerate}
\usepackage{colortbl}

\urldef{\mailsa}\path|dyyslfan@gmail.com, yangyubin@nju.edu.cn|

\newcommand{\keywords}[1]{\par\addvspace\baselineskip

\renewcommand\floatpagefraction{.9}
\renewcommand\topfraction{.9}
\renewcommand\bottomfraction{.9}
\renewcommand\textfraction{.1}
\setcounter{totalnumber}{50}
\setcounter{topnumber}{50}
\setcounter{bottomnumber}{50}
\noindent\keywordname\enspace\ignorespaces#1}
\newcommand{\RNum}[1]{\uppercase\expandafter{\romannumeral #1\relax}}

\begin{document}

\mainmatter

\title{Efficient Web Service Composition via Knapsack-Variant Algorithm}

\author{Shiliang Fan$^1$
\and Yubin Yang$^{1,*}$}

\institute{$^1$State Key Laboratory for Novel Software Technology, \\
Nanjing University, Nanjing 210023, China\\
\mailsa\\}

\maketitle

\begin{abstract}
Since the birth of web service composition, minimizing the number of web services of the resulting composition while satisfying the user request has been a significant perspective of research. With the increase of the number of services released across the Internet, seeking efficient algorithms for this research is an urgent need. In this paper we present an efficient mechanism to solve the problem of web service composition. For the given request, a service dependency graph is firstly generated with the relevant services picked from an external repository. Then, each search step on the graph is transformed into a dynamic knapsack problem by mapping services to items whose volume and cost is changeable, after which a knapsack-variant algorithm is applied to solve each problem after transformation. Once the last search step is completed, the minimal composition that satisfies the request can be obtained. Experiments on eight public datasets proposed for the Web Service Challenge 2008 shows that the proposed mechanism outperforms the state-of-the-art ones by generating solutions containing the same or smaller number of services with much higher efficiency.
\keywords{Web Service Composition, Minima, Efficient, Knapsack.}
\end{abstract}

\section{Introduction}

Web services are platform-independent applications which are released, discovered and invoked over the web using open standards such as UDDI \cite{ref1}, SOAP \cite{ref2}, and WSDL \cite{ref3}. As software modules published on servers and consumed across the Internet, web services transmit their communication data over the network expediently, which leads to the loosely-coupled nature. The nature of loose coupling pave the way for easier and wider-ranging integration and interoperability among systems, making the technology of web services extensively used in enterprises. It is as plain as a pikestaff that web services have been the support technology for the swift development of IT-based services economy.

However, the complex business requirements cannot be fulfilled by a single service in most cases. But meanwhile, with the sharp increase of the number of web services, the composition of web services provides a way to solve the problem. Web service composition is the process of building a more complex, functional workflow via combining a collection of single services together, to satisfy the inputs and outputs given by users. There are mainly two kinds of approaches for the web service composition problem. Some researches transform the composition problem into a planning one by mapping services to actions \cite{ref4}, \cite{ref5}, \cite{ref6}, which is known as AI-based technique. Others construct a graph to express the relationship of services and aim to extract a reachable path from the graph \cite{ref7}, \cite{ref8}, which is called graph-based technique.

There exists a phenomenon that a growing number of services own similar or identical functionality. Therefore, it is simple to known that, for a given request, the composition process of massive services may generate many possible solutions with different number of services. Minimizing the number of services of the resulting composition while satisfying the user request is significant for brokers, customers, and providers \cite{ref9}. Standing in the shoes of brokers, a composition result with fewer services could make it easier for the work of maintenance and management. From the customers' point of view, a smaller composition ordinarily means the less payment for those services invoked; On the other side, decreasing the number of services involved in the composition may highly increase the success rate of acquiring the wanted responses to the requests made by customers. As for service providers, solutions with fewer services could save resources and cost for the same task.

Up to now, there are several studies on the web service composition taking the optimization of the number of services into consideration. Nevertheless, in face of the repositories that contains a substantial amount of services, existing researches take too long time to obtain the optimal solution on account of the huge search space. Thus these approaches aren't efficient enough to be applied to large-scale and real-time environments. In this paper, we aim at presenting a mechanism to efficiently solve the web service composition problem. The main contribution are:
\begin{itemize}
\item[$\bullet$] An equivalent transformation approach that transforms search steps on the service dependency graph into dynamic knapsack problems by mapping services to items with changeable volume and cost.
\item[$\bullet$] A knapsack-variant algorithm that guarantees to efficiently generate the composition with minimal number of services by means of solving each dynamic knapsack problem.
\item[$\bullet$] An optimization strategy to reduce the spatial complexity of the knapsack-variant algorithm.
\end{itemize}
Furthermore, a full validation on eight datasets of Web Service Challenge 2008 has been done. Experimental results show that our mechanism performs better than the state-of-the-arts both in term of quality and efficiency.

The rest of this paper is organized as follows. Section 2 reviews some related work, Section 3 describes the background and formalizes the web service composition problem, then illustrates the motivation of this research. Section 4 introduces the proposed mechanism, Section 5 shows the experimental results, and Section 6 provides some final remarks.

\section{Related Work}

Effectively combining minimal number of services distributed over the web to build enterprise-class services that satisfy given requirements is the goal of this paper. A survey of the problem of web service composition shows that several researches have been done in this perspective, and each has its own merits.

A heuristic A* search algorithm is proposed in \cite{ref7} for the problem of automatic web service composition. For a given request, a digraph called service dependency graph is constructed firstly with a part of the original services chosen from an external repository. Meanwhile, some techniques are applied to reduce the useless nodes in the graph. Then the heuristic-based search algorithm named A* is executed over the optimized graph to seek the minimal composition which fulfills the user request. Though it can obtain compositions with minimal number of services on WSC-2008's datasets, it may show a poor performance in large-scale and  real-time environments. On one hand, different kinds of optimizations on the service dependency graph may spend large quantities of time. On the other hand, quite a few iterations of A* search algorithm aren't applicable to real-time scenarios.

A scalable and approximate mechanism is presented in \cite{ref10} to get the near-minimal compositions against time. The authors proposed an on-the-fly strategy to construct only a path of the auxiliary graph instead of the complete graph. Additionally, a deterministic approach and a probabilistic approach are discussed to find the path with the minimal number of services, which is the final result of composition. Though the algorithm has a superior service composition time compared to other algorithms, the greedy strategy adopted always gets stuck in local optima. As a result, it always generates compositions with more services than the others. In a word, it performs well in efficiency but it remains to be improved in terms of the quality of solutions.

Pablo et al. \cite{ref11} present a composition framework integrating fine-grained I/O service discovery strategy and an optimal composition search algorithm. To improve the efficiency of the generation of a layered service composition graph, the discovery and matchmaking phases are optimized using indexes and cache. Once the graph is generated, many optimizations are applied to reduce the graph size. Then a search which is modelled as a state-transition system is performed over the graph to find the minimal composition among all the possible compositions satisfying the request provided by user. Experimental results show the scalability and flexibility of the composition framework. However, similar with the mechanism in \cite{ref7}, though lots of optimizations are used to improve the optimal composition search performance, much extra time is spent in the step. On the other side, the search algorithm isn't efficient enough to be applied to large-scale and real-time environments.

In summary, despite above algorithms to optimize number of services, there is a lack of approaches that have the ability to minimize the number of services of the composition effectively and efficiently. This paper proposes an effective and efficient mechanism in an effort to find compositions with minimal number of services in large-scale and real-time scenarios.

\section{Preliminaries and Motivation}

\subsection{Preliminary Knowledge}

Web service composition is a well studied problem, and semantic web services are the foundation of the problem. In this paper, a {\it semantic web service} is formally defined as follows \cite{ref12}, \cite{ref13}, \cite{ref14}.

\begin{definition}
Given a set of concepts named Con, a Semantic Web Service ("service" for short) is defined as a tuple s = \{$In_s, Out_s$\}, where $In_s$ = \{$in^1_s, \ldots, in^n_s$\} is the set of inputs required to invoke the semantic web service s, and $Out_s$ = \{$out^1_s, \ldots, out^n_s$\} is the set of outputs generated by executing the service s. Each element of $In_s$ and $Out_s$ is actually a semantic concept belonging to the set Con, namely, $In_s$ $\subseteq$ Con and $Out_s$ $\subseteq$ Con.
\end{definition}

Individual services can be combined by connecting their matched inputs and outputs to construct compositions \cite{ref15}, \cite{ref16}.

\begin{lemma}
Given an output $out_s$ $\in$ $Out_s$ of a service s, as well as an input $in_{s^\prime}$ $\in$ $In_{s^\prime}$ of another service $s^\prime$, if $out_s$ and $in_{s^\prime}$ are equivalent concepts or $out_s$ is a sub-concept of $in_{s^\prime}$, $out_s$ matches $in_{s^\prime}$ (i.e., $in_{s^\prime}$ is matched by $out_s$).
\end{lemma}

There are mainly two kinds of structures of these compositions, namely sequential structure and parallel structure \cite{ref17}. The services organized as a sequential structure are invoked in order, while the services organized as a parallel structure are invoked synchronously. A {\it composition} can be described as follows.

\begin{definition}
A Composition containing the set of services $S = \{s_1, \ldots, s_n\}$ is defined as $\Omega_S = s_1, \ldots, s_n$. If the services are chained in sequence, the composition is expressed as $\Omega_S^\rightarrow = s_1$$\rightarrow$$\ldots$$\rightarrow$$s_n$; if in parallel, then $\Omega_S^{\parallel} = s_1$$\parallel$$\ldots$$\parallel$$s_n$. The set of services involved in $\Omega_S$ is defined as $Servs(\Omega_S) = S$. Moreover, the length of a composition $\Omega_S$ is defined as $Len(\Omega_S) = \vert S \vert$, namely the number of services in  $\Omega_S$.
\end{definition}

The aim of the service composition problem is to automatically select the minimal composition of available services to fulfil a user request that is defined as follows.

\begin{definition}
A Request from users is defined as a tuple R = \{$In_R, Out_R$\}, where $In_R$ = \{$in^1_R, \ldots, in^n_R$\} is the set of inputs provided by users ($In_R$ $\subseteq$ Con), and another element $Out_R$ = \{$out^1_R, \ldots, out^n_R$\} represents the set of expected outputs ($Out_R$ $\subseteq$ Con).
\end{definition}

On the basis of the above concepts, the precise definition of the {\it web service composition problem} is given as follows.

\begin{definition}
A Web Service Composition Problem is defined as, for a given composition request $R$, to seek for a composition $\Omega_S$ fulfilling $R$ with the optimization objective of $\bf{min}$$\mid$\,$S$\,$\mid$, namely, the $\Omega_S$ contains the minimal number of services.
\end{definition}

\subsection{A Motivating Example}

Graph is a natural and intuitive way to express the complex interaction relations between entities. {\it Service Dependency Graph} is a digraph used to describe services and the matching relations among them \cite{ref8}, \cite{ref10}, \cite{ref12}. For a given request $R = \{\{in_1, in_2\}, \{out_1, out_2\}\}$, an example of service dependency graph is shown in Fig.1. Each service is represented as a rectangle, while the inputs and outputs of a service are represented as circles. Furthermore, the matching relations among services are represented as edges connecting two circles.

\begin{figure}[h]
\centering
\includegraphics[scale=0.34]{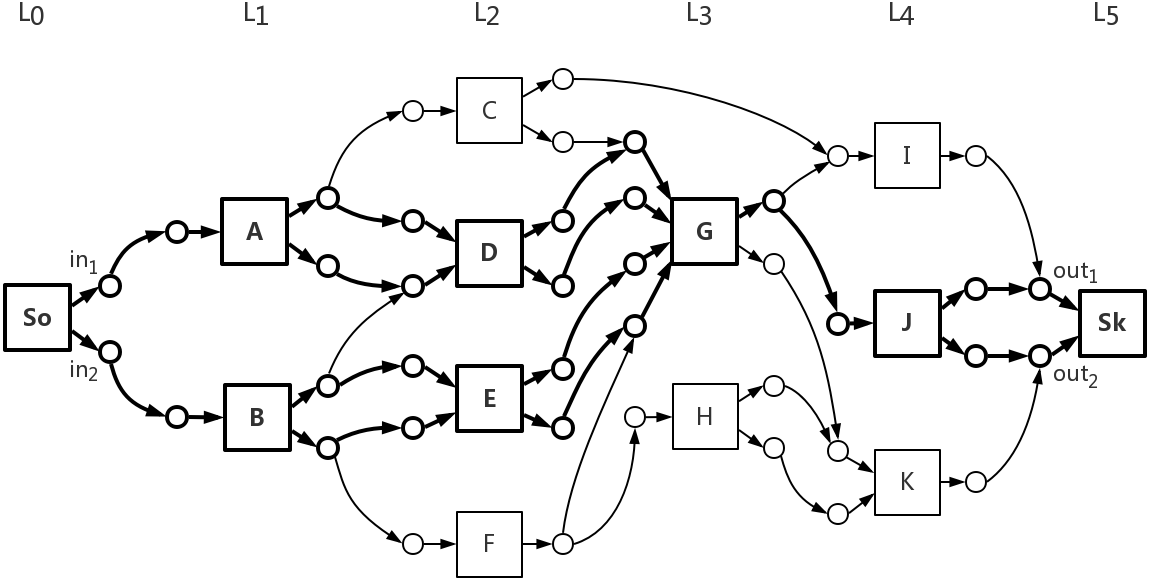}
\caption{A service dependency graph example with the optimal composition highlighted.}
\end{figure}

As can be seen from Fig.1, the highlighted composition which can be represented as $\Omega = s_o$$\rightarrow$$((A$$\rightarrow$$D)$$\parallel$$(B$$\rightarrow$$E))$$\rightarrow$$G$$\rightarrow$$J$$\rightarrow$$s_k$ contains eight services in total (including the $s_o$ and the $s_k$). There are also several other compositions satisfying the same user request $R$ such as $\Omega^\prime = s_o$$\rightarrow$$((A$$\rightarrow$$C$$\rightarrow$$I)$$\parallel$$(B$$\rightarrow$$F$$\rightarrow$$H$$\rightarrow$$K))$$\rightarrow$$s_k$, whereas the number of services of them are unexceptionally more than eight. Therefore, the composition $\Omega$ highlighted in the graph is the optimal one with the minimal number of services.

In large-scale scenarios, the service dependency graph may be exceedingly complex, which leads to a huge search space \cite{ref18}. As a consequence, it is formidable to extract the optimal composition from the graph. There is no doubt that the exhaustive combinatorial search can guarantee the optima, but it will take an unacceptable long time to generate the compositions and isn't applicable to real-time environments. To sum up, we should pay attention not only to the quality of the resulting composition but also to the efficiency of the composition algorithm.

\section{Detailed Methodology}

In this section, an efficient mechanism is proposed for the problem of web service composition. Given a composition request $R = \{In_R, Out_R\}$ and a service repository $S_r$, a service dependency graph is firstly constructed with the relevant services for the request. Then, search steps on the graph are transformed into dynamic knapsack problems, after which a knapsack-variant algorithm is proposed to solve each dynamic knapsack problem. Finally, an optimization strategy is adopted to reduce the spatial complexity of the knapsack-variant algorithm.

\subsection{Service Dependency Graph}

As shown in Fig.1, a service dependency graph is a layered digraph. The first layer contains only one dummy service $s_o = \{\varnothing, In_R\}$, similarly, there is only a dummy service $s_k = \{Out_R, \varnothing\}$ in the last layer, while the concrete services in the other layers are selected from $S_r$. Moreover, each layer contains the services whose inputs are all matched by the outputs generated by previous layers.

The generation process of a service dependency graph is shown in Algorithm 1. Given the request $R$ and the repository $S_r$, $s_o$ is firstly added to the first layer $L_0$, after which each following layer $L_i$ is constructed with the services whose inputs are all matched by the outputs generated by previous layers. $s_k$ will be added to the last layer if the set of expected outputs $Out_R$ is included in $Out_{all}$. Finally, unused services making no contribution to $Out_R$ are removed from the graph by traversing from the last layer to the first layer.

\begin{algorithm}[h]
  \small
  \caption{Construction of Service Dependency Graph}
  \KwIn{$R$, $S_r$}
  \KwOut{$L$}
  $i$ $\gets$ $0$, $L_i$ $\gets$ $\{s_o\}$, $i$ $\gets$ $i+1$, $Out_{all}$ $\gets$ $In_R$ \\
  \Repeat{$Out_R \subseteq Out_{all}$}
  {
      \For{{\rm service} $s \in S_r$}
      {
          \If{$s \notin L_j(\forall j<i)$ \ {\rm and} \ $In_s \subseteq Out_{all}$}
          {
              $L_i$ $\gets$ $L_i \cup \{s\}$ \\
              $Out_{all}$ $\gets$ $Out_{all} \cup Out_s$
          }
      }
      $i$ $\gets$ $i+1$
  }
  $tot$ $\gets$ $i$, $L_{tot}$ $\gets$ $\{s_k\}$, $j$ $\gets$ $tot$, $In_{all}$ $\gets$ $Out_R$ \\
  \While{$j \ge 0$}
  {
      \For{{\rm service} $s \in L_j$}
      {
          \If{$Out_s \cap In_{all} = \varnothing$}
          {
              $L_j$ $\gets$ $L_j - \{s\}$ \\
          }
      }
      \For{{\rm service} $s \in L_j$}
      {
          $In_{all}$ $\gets$ $In_{all} \cup In_s$ \\
      }
      $j$ $\gets$ $j - 1$
  }
  \Return $L$
\end{algorithm}

\subsection{The Dynamic Knapsack Problem}

Once the service dependency graph is completed, the composition problem is regarded as searching for a reachable path from $s_o$ to $s_k$. Each search step on the graph is defined as determining the optimal {\it precursors} of each service.
\begin{definition}
The set of precursors of a service $s \in L_i$ is defined as $Pre(s) = \{s^\prime \, \vert \, s^\prime \in L_j(\forall j<i) \land In_s \cap Out_{s^\prime} \not= \varnothing\}$. Specially, Pre($s_o$) = $\varnothing$.
\end{definition}

The search step of service $G$ is shown in Fig.2. Note that, the minimal composition ending with a service $s$ is expressed as $\Omega^s$, and $c_i$ in the figure represents an input or output concept of services. Assuming that the minimal compositions ending with the precursors of $G$, i.e., $\Omega^C$, $\Omega^D$, $\Omega^E$, and $\Omega^F$, have been determined in advance, the search step of $G$ is defined as selecting the optimal subset of $\{\Omega^C, \Omega^D, \Omega^E, \Omega^F\}$ to compose the $\Omega^G$, which is actually a greedy strategy. On the basis of the greedy strategy, an equivalent transformation approach is proposed to transform each search step similar to the one shown in Fig.2 into a {\bf Dynamic Knapsack Problem}.

\begin{figure}[h]
\begin{minipage}[t]{0.5\linewidth}
\centering
\includegraphics[width=1.95in]{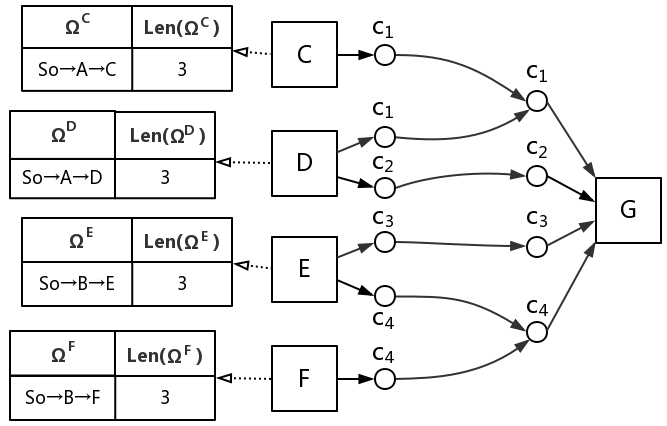}
\caption{A search step on the graph.}
\label{fig:side:a}
\end{minipage}%
\begin{minipage}[t]{0.5\linewidth}
\centering
\includegraphics[width=1.95in]{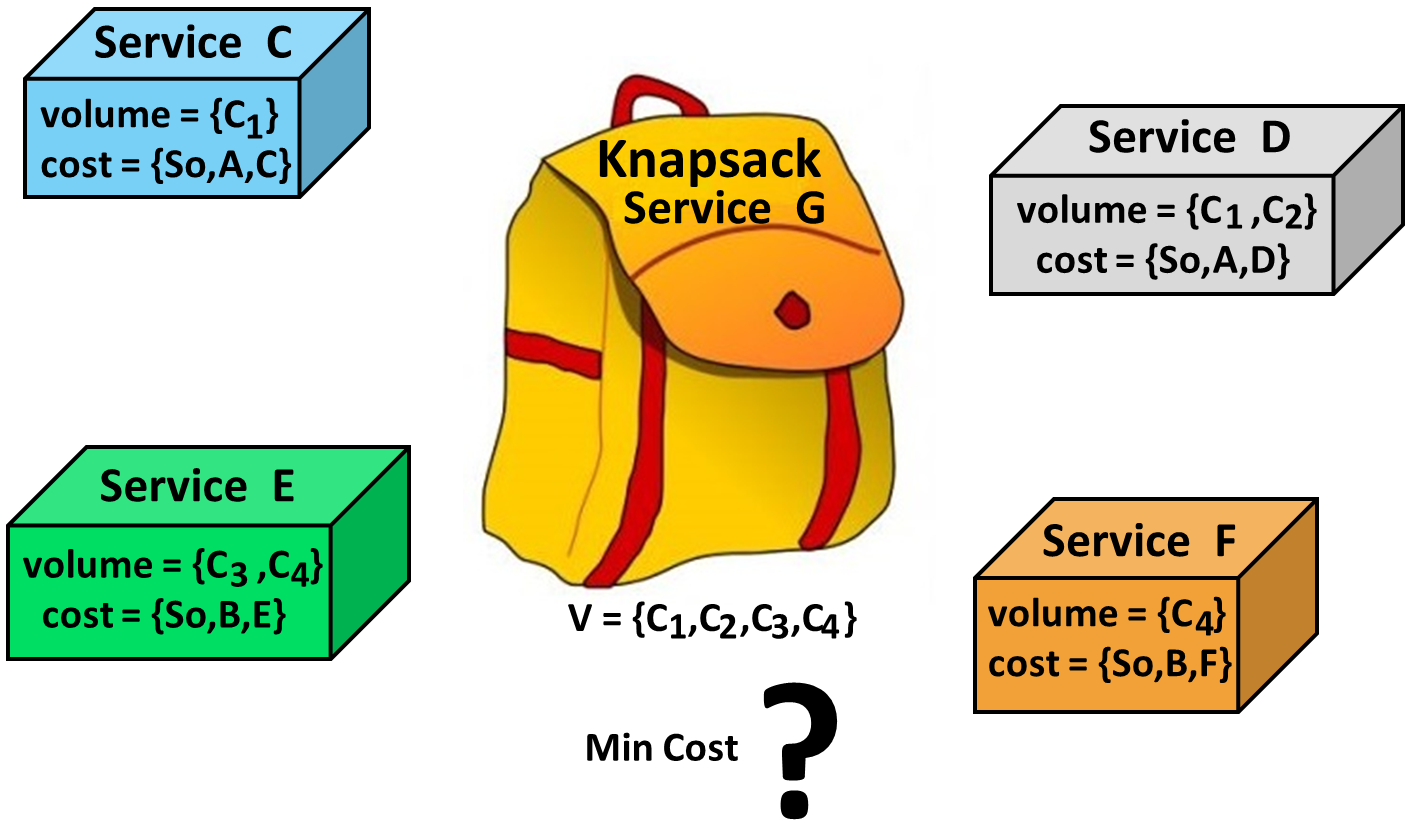}
\caption{The dynamic knapsack problem.}
\label{fig:side:b}
\end{minipage}
\end{figure}

As shown in Fig.3, assuming that the optimal precursors of a service $s$ require to be determined, $s$ is abstracted into a knapsack whose capacity is $In_s$ (the set of inputs of the service $s$). Each precursors of $s$ is spontaneously regarded as an item with {\bf dynamic} {\it volume} and {\it cost}. The problem is to minimize the sum of the cost of the items in the knapsack so that the sum of the volume is equal to the knapsack's capacity. The volume of an item $s^\prime \in Pre(s)$ is relevant to $Out_{s^\prime}$ (the set of outputs of the service $s^\prime$), and the cost of $s^\prime$ is measured with $Servs(\Omega^{s^\prime})$ (the set of services involved in the composition $\Omega^{s^\prime}$), which is too inconvenient to be applied to the following composition algorithm. Therefore, an approach is presented to quantify the volume and the cost of each item.

\begin{algorithm}[h]
  \small
  \caption{Generation of Subsets}
  \KwIn{$V$}
  \KwOut{$Subs$}
  $Subs$ $\gets$ $\{\varnothing\}$, $upper\_bound$ $\gets$ $2^{\vert V \vert}$ \\
  \For{$index = 0; index < upper\_bound; index$$+$$+$}
  {
      $i$ $\gets$ $0$, $tmp$ $\gets$ $index$, $subset$ $\gets$ $\{\}$ \\
      \While{$tmp > 0$}
      {
          \If{$(tmp$ {\rm mod} $2)$ $>$ $0$}
          {
              $subset$ $\gets$ $subset \cup \{V[i]\}$
          }
          $tmp$ $\gets$ $tmp$ {\rm div} $2$, $i$ $\gets$ $i + 1$\\
      }
      $Subs[index]$ $\gets$ $subset$ \\
  }
  \Return $Subs$
\end{algorithm}

\subsubsection*{The quantization of volume.}

Firstly, all the subsets of $In_s$ is obtained by Algorithm 2 in a certain order. Then, on the ground of the returned subsets $Subs$, a mapping table is constructed to quantify the volume of the knapsack and each item. Taking the problem in Fig.3 as an example, the mapping table used to support the volume quantization is shown as Table 1.

\begin{table}[h]
  \fontsize{8}{8}\selectfont
  \centering
  \begin{threeparttable}
  \tabcolsep0.7pt
  \caption{The mapping table.}
    \begin{tabular}{ccccccccccccccccc}
    \toprule[1.2pt]
    \multicolumn{1}{c}{{\bf index}}&{{0}}&{{1}}&{{2}}&{{3}}&{{4}}&{{5}}&{{6}}&{{7}}\cr
    \midrule[0.6pt]
    \multicolumn{1}{c}{Subs[index]}&$\varnothing$&$\bf \{c_1\}$&$\{c_2\}$&$\bf \{c_1,c_2\}$&$\{c_3\}$&$\{c_1,c_3\}$&$\{c_2,c_3\}$&$\{c_1,c_2,c_3\}$\cr
    \midrule[0.6pt]
    \multirow{1}*{item/knapsack}&{}&{\bf C}&{}&{\bf D}&{}&{}&{}&{}\cr
    \midrule[0.6pt]
    \multicolumn{1}{c}{volume}&0&{\bf 1}&2&{\bf 3}&4&5&6&7\cr
    \midrule[0.6pt]
    \multicolumn{1}{c}{index(binary)}&0000&{\bf 0001}&0010&{\bf 0011}&0100&0101&0110&0111\cr
    \toprule[1.2pt]
    \multicolumn{1}{c}{{\bf index}}&{{8}}&{{9}}&{{10}}&{{11}}&{{12}}&{{13}}&{{14}}&{{15}}\cr
    \midrule[0.6pt]
    \multicolumn{1}{c}{Subs[index]}&$\bf \{c_4\}$&$\{c_1,c_4\}$&$\{c_2,c_4\}$&$\{c_1,c_2,c_4\}$&$\bf \{c_3,c_4\}$&$\{c_1,c_3,c_4\}$&$\{c_2,c_3,c_4\}$&$\bf \{c_1,c_2,c_3,c_4\}$\cr
    \midrule[0.6pt]
    \multirow{1}*{item/knapsack}&{\bf F}&{}&{}&{}&{\bf E}&{}&{}&{\bf G}\cr
    \midrule[0.6pt]
    \multicolumn{1}{c}{volume}&{\bf 8}&9&10&11&{\bf 12}&13&14&{\bf 15}\cr
    \midrule[0.6pt]
    \multicolumn{1}{c}{index(binary)}&{\bf 1000}&1001&1010&1011&{\bf 1100}&1101&1110&{\bf 1111}\cr
    \bottomrule[1.2pt]
    \end{tabular}
    \end{threeparttable}
\end{table}

By means of Table 1, the volume of the knapsack and each item can be quantified as follows.
\begin{itemize}
\item[$\circ$] The capacity of the knapsack $s$ is quantified as the upper bound of $index$, namely $\vert Subs \vert - 1$.
\item[$\circ$] Assuming that the service $s^\prime$ provides the set of outputs $Out \subseteq Out_{s^\prime}$ for the service $s$, the volume of $s^\prime$ is quantified as the value of the $index$ which satisfies the condition that $Subs[index] = Out$.
\end{itemize}
Taking the problem in Fig.3 as an instance, the capacity of the knapsack $G$ is $V_{cap} = 15$ after quantization. Two different feasible solutions of the dynamic knapsack problem are shown in Fig.4 respectively. As can be seen from the solution \RNum{1}, the {\it Service D} provides the set of outputs $\{c_1, c_2\}$ for the {\it Service G}, thus the volume of the item $D$ is quantified as $volume_D = 3$ according to Table 1. In addition, $volume_E = 4$ owing to the fact that the {\it Service E} provides the set of outputs $\{c_3\}$ for {\it G}. Similarly, $volume_F = 8$. It is not difficult to observe that $volume_D + volume_E + volume_F = 3 + 4 + 8 = V_{cap}$, hence the knapsack $G$ can be filled with the set of items $\{D, E, F\}$, which indicates the effectiveness of the quantization.

The volume of an item is changeless in the 0-1 knapsack problem, while the volume of an item is changeable in the dynamic knapsack problem. For example, let's discuss the solution \RNum{2} shown in Fig.4(b). Despite the fact that the set of outputs of the {\it Service D} is $\{c_1, c_2\}$, $D$ only provides $\{c_2\}$ for $G$ seeing that $\{c_1\}$ is provided by the {\it Service C}, which leads to the change of $volume_D$ from 3 to 2. Meanwhile, $volume_C + volume_D + volume_E + volume_F = 1 + 2 + 4 + 8 = V_{cap}$. The outputs provided by the service $s^\prime$ for the service $s$ are uncertain before decision-making, hence the volume of the item $s^\prime$ can't be determined in advance.

\begin{figure}[h]
  \centering
  \subfigure[The solution \RNum{1}]{
    \label{fig:subfig:a} %% label for first subfigure
    \includegraphics[width=1.98in]{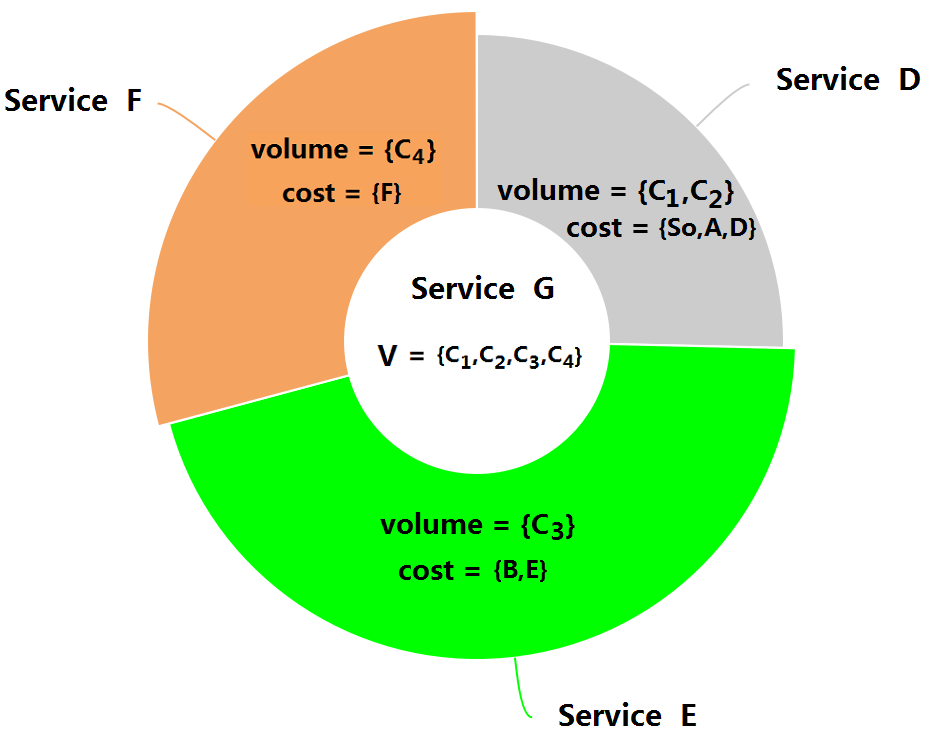}}
  \hspace{0.0cm}
  \subfigure[The solution \RNum{2}]{
    \label{fig:subfig:b} %% label for second subfigure
    \includegraphics[width=2.2in]{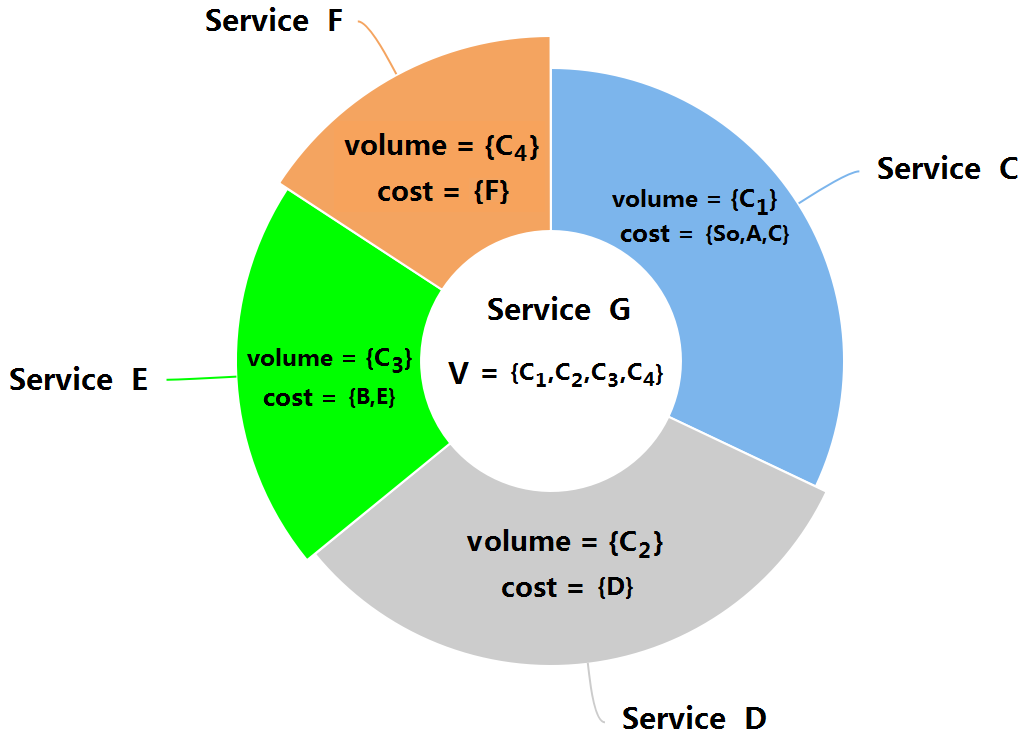}}
  \caption{Two different feasible solutions to fill $G$.}
  \vspace{-0.5cm}
  \label{fig:subfig} %% label for entire figure
\end{figure}

\subsubsection*{The quantization of cost.}

Considering that the goal of this paper is minimizing the number of services involved in the resulting composition, the cost of an item $s^\prime$ is designed as the minimal number of services that require to be invoked to generate the outputs of the service $s^\prime$. Therefore, the cost of each item is quantified as follows.
\begin{itemize}
\item[$\circ$] Assuming that $Ser$ represents the set of services which belong to $Servs(\Omega^{s^\prime})$ and haven't been invoked yet, the volume of the item $s^\prime$ is quantified as the size of the $Ser \cup \{s^\prime\}$.
\end{itemize}
Taking the solution \RNum{1} shown in Fig.4(a) as an example, suppose that the set of items $\{D, E, F\}$ are put into the knapsack $G$ in order of the name ($D$$\rightarrow$$E$$\rightarrow$$F$). We have already assumed that the minimal compositions ending with the precursors of $G$ have been determined in advance. Figure 2 reveals that $\Omega^D = $$s_o$$\rightarrow$$A$$\rightarrow$$D$, $\Omega^E = $$s_o$$\rightarrow$$B$$\rightarrow$$E$ and $\Omega^F = $$s_o$$\rightarrow$$B$$\rightarrow$$F$. On account of the fact that $D$ is included in the knapsack $G$ before $E$, the set of services $\{s_o\}$ has been invoked in advance. As a result, only $\{B\}$ requires to be invoked before $E$, which leads to $cost_E = 2$. Similarly, $cost_D = 3$, $cost_F = 1$ and the total cost of the solution \RNum{1} is calculated as $cost_D + cost_E + cost_F = 3 + 2 + 1 = 6$.

The cost of an item is changeable as well in the dynamic knapsack problem. As shown in the solution \RNum{2} (assume that the items are put into the knapsack $G$ in order of $C$$\rightarrow$$D$$\rightarrow$$E$$\rightarrow$$F$), the set of services $\{s_o, A\}$ has been invoked when putting the item $C$ into the knapsack. Consequently $D$ can be invoked directly, which leads to the change of $cost_D$ from 3 to 1.

In summary, only by ingeniously determining the dynamic volume and cost of an item can we solve the dynamic knapsack problem.

\subsection{The Knapsack-Variant Algorithm}

The dynamic knapsack problem aiming at determining the optimal precursors of a service $s$ can be described as follows. Given a knapsack $s$ whose volume capacity is $V_{cap}$, and a set of items $Pre(s) = \{s_1, s_2, \dots, s_N\}$ where $N = \vert Pre(s)\vert$ represents the number of items, each with a dynamic volume $volume_i$ and a dynamic cost $cost_i$, some items are selected from $Pre(s)$ to fill the knapsack $s$ with the objective of:

\begin{equation}
\begin{aligned}
& \bf{minimize}
& & \sum\limits_{i=1}^{N}{cost_i \cdot x_i} \\
& \text{subject to}
& & \sum\limits_{i=1}^{N}{volume_i \cdot x_i}{ \ = \ V_{cap}}, \\
&&& \ x_i \in \{0,1\}.
\end{aligned}
\end{equation}
\noindent where $x_i$ represents the number of the item $i$ to include in the knapsack. Unlike the 0-1 knapsack problem, all the $volume_i$ and $cost_i$ are uncertain here, which leads to the inapplicability of the {\it 0-1 Knapsack Algorithm}. In this section, a {\bf Knapsack-Variant Algorithm} is proposed to solve the problem by determining the volume and cost of each service dynamically.

Let $C[i][v]$ represent the minimal cost of selecting items from $\{s_1, s_2, \dots, s_i\}$ ($1\le i\le N$) to fill a temporary knapsack whose capacity is $v$ ($1\le v\le V_{cap}$), and $I[i][v]$ the set of items selected to minimize $C[i][v]$. Then,
\begin{equation}
\begin{aligned}
& C[i][v] \, \, \, =
& & {\bf min} \ \{C[i-1][v], C[i-1][v-volume_i]+cost_i\} \\
& \text{where}
& & {volume_i = {\bf DV}}{(s_i, In_s, Subs, v)}, \\
&&& {cost_i = {\bf DC}}{(s_i, I, i, v ,volume_i)}.
\end{aligned}
\end{equation}
$volume_i$ and $cost_i$ are determined dynamically in the process of the dynamic programming, which is the quintessence of the knapsack-variant algorithm.

The function $DV$ in Algorithm 3 is used to dynamically calculate the volume of an item. For the given temporary knapsack with the capacity of $v$, the outputs provided by service $s_i$ for the knapsack are determined as $Out_{s_i} \cap Subs[v]$. Thus, the volume of the item $s_i$ can be quantified by the quantization approach proposed in Sect.4.2. Inspired by the one-to-one match between the {\it index(binary)} and the {\it Subs[index]} in Table 1, a binary method is applied to determine the $index$ which satisfies $Subs[index] = Out_{s_i} \cap Subs[v]$, namely the $volume_i$.

\begin{algorithm}[h]
  \small
  \caption{Determination of Volume of Items}
  \KwIn{$s_i$, $In_s$, $Subs$, $v$}
  \KwOut{$DV(s_i, In_s, Subs, v)$}
  $map$ $\gets$ $\{\}$,  $volume$ $\gets$ $0$, $Out$ $\gets$ $Out_{s_i} \cap Subs[v]$ \\
  \For{$index = 0; index < \vert In_s \vert; index$$+$$+$}
  {
      $c$ $\gets$ $In_s[index]$ \\
      $map[c]$ $\gets$ $index$
  }
  \For{\rm concept $c \in Out$}
  {
      $index$ $\gets$ $map[c]$ \\
      $volume$ $\gets$ $volume$ + $2^{index}$
  }
  $DV(s_i, In_s, Subs, v)$ $\gets$ $volume$ \\
  \Return $DV(s_i, In_s, Subs, v)$
\end{algorithm}

For the given temporary knapsack with the capacity of $v$, the reason why the outputs provided by $s_i$ for the knapsack are determined as $Out_{s_i} \cap Subs[v]$ can be explained as follows. The knapsack with the capacity of $v$ corresponds to a temporary service $s_v$ with the inputs of $Subs[v]$. Therefore, the largest set of outputs provided by the service $s_i$ for $s_v$ is obviously $Out_{lar} = Out_{s_i} \cap Subs[v]$. However, if a smaller one $Out_{sma} \subset Out_{lar}$ is provided for $s_v$, $Out_{lar} - Out_{sma}$ may require to be provided by extra services selected from $\{s_1, s_2, \dots, s_{i-1}\}$, which causes the loss of the local optimum, let alone the global optimum.

Moreover, the function $DC$ shown in Algorithm 4 is applied to determine the cost of an item $s_i$. Since the items that have been included in the knapsack are cached in the data structure $I$, the set of services $Ser \subseteq Servs(\Omega^{s_i})$ which haven't been invoked can be obtained drawing support from $I$, after which the cost of the item $s_i$ is quantified as $\vert Ser\vert + 1$.

\begin{algorithm}[h]
  \small
  \caption{Determination of Cost of Items}
  \KwIn{$s_i$, $I$, $i$, $v$, $volume_i$}
  \KwOut{$DC(s_i, I, i, v, volume_i)$}
  $Union$ $\gets$ $\{\}$ \\
  \For{{\rm service} $s \in I[i-1][v-volume_i]$}
  {
      $Union$ $\gets$ $Union \cup Servs(\Omega^s)$
  }
  $Inter$ $\gets$ $Servs(\Omega^{s_i}) \cap Union$ \\
  $Ser$ $\gets$ $Servs(\Omega^{s_i}) - Inter$ \\
  $DC(s_i, I, i, v, volume_i)$ $\gets$ $\vert Ser\vert + 1$ \\
  \Return $DC(s_i, I, i, v, volume_i)$
\end{algorithm}

According to the optimization model in (2), by systematically increasing the values of $i$ (from 1 to $N$) and $v$ (from 1 to $V_{cap}$), the composition $\Omega^s$ with the minimal number of services will be finally obtained when $i = N$ and $v = V_{cap}$.
\begin{equation}
Len(\Omega^s) = C[N][V_{cap}] + 1.
\end{equation}
Therefore, the time complexity of the search step is $O(NV_{cap})$, so is the spatial complexity. However, the spatial complexity of (2) can be further optimized.

Considering that $C[i][v]$ is only relevant to $C[i-1][v^\prime]$ ($1\le v^\prime\le v$), $C[i][v]$ can be replaced by an one-dimensional array $C[v]$. Then,
\begin{equation}
\begin{aligned}
& C[v] \, \, \, =
& & {\bf min} \ \{C[v], C[v-volume_i]+cost_i\} \\
& \text{where}
& & {volume_i = {DV}}{(s_i, In_s, Subs, v)}, \\
&&& {cost_i = {DC}}{(s_i, I, i, v ,volume_i)}.
\end{aligned}
\end{equation}
The problem can be solved by systematically increasing the values of $i$ (from 1 to $N$) and decreasing $v$ (from $V_{cap}$ to 1), hence the spatial complexity is reduced from $O(NV_{cap})$ to $O(V_{cap})$.

The knapsack-variant algorithm is shown in Algorithm 5. Search steps on the graph are carried out layer by layer. Each search step depends on the optimization results of search steps in previous layers and is transformed into a knapsack problem that can be solved by (4). After completing the last search step of $s_k$, the expected composition with the length of $Len(\Omega^{s_k})$ can be obtained.

\begin{algorithm}[t]
  \small
  \caption{Knapsack-Variant Algorithm}
  \KwIn{$L$}
  \KwOut{$Len(\Omega^{s_k})$}
  $Servs(\Omega^{s_o})$ $\gets$ $\{s_o\}$, $Len(\Omega^{s_o})$ $\gets$ $1$ \\
  \For{$index = 1; index < \vert L \vert; index$$+$$+$}
  {
      \For{{\rm service} $s \in {L}_{index}$}
      {
          $pres$ $\gets$ $Pre(s)$, $N$ $\gets$ $\vert pres \vert$ \\
          $Subs$ $\gets$ all subsets of $In_s$, $V_{cap}$ $\gets$ $\vert Subs \vert - 1$\\
          $C[0..V_{cap}]$ $\gets$ $+\infty$, $C[0]$ $\gets$ $0$, $I[0..N][0..V_{cap}]$ $\gets$ $\{\}$ \\
          \For{$i = 1; i <= N; i$$+$$+$}
          {
              $s_i$ $\gets$ $pres[i]$ \\
              \For{$v = V_{cap}; v > 0; v$$-$$-$}
              {
                  $volume_i$ $\gets$ $DV(s_i, In_s, Subs, v)$ \\
                  $cost_i$ $\gets$ $DC(s_i, I, i, v, volume_i)$ \\
                  \If{$C[v-volume_i] + cost_i < C[v]$}
                  {
                      $C[v]$ $\gets$ $C[v-volume_i] + cost_i$ \\
                      $I[i][v]$ $\gets$ $I[i-1][v-volume_i] \cup {s_i}$
                  }
                  \Else
                  {
                      $I[i][v]$ $\gets$ $I[i-1][v]$
                  }
              }
          }
          $Servs(\Omega^s)$ $\gets$ $\{\}$ \\
          \For{{\rm service} $item \in I[N][V_{cap}]$} {
              $Servs(\Omega^s)$ $\gets$ $Servs(\Omega^s) \cup Servs(\Omega^{item})$
          }
          $Len(\Omega^s)$ $\gets$ $C[V_{cap}] + 1$ \\
      }
  }
  \Return $Len(\Omega^{s_k})$
\end{algorithm}

\section{Experimental Results}

In order to evaluate the performance of the proposed composition algorithm, we completed a group of experiments on eight public repositories from the 2008 Web Service Challenge. Services in each repository are defined using a WSDL file, and inputs and outputs are semantically described in a XML file called ontology.

Table 1 shows the detailed characteristics of each dataset. The number of services and concepts in the ontology of each dataset are shown in rows {\it \#Services} and {\it \#Concepts} respectively. Row {\it \#Sol.Services} indicates the number of services for the optimal solution provided by the WSC'08.
\begin{table}[h]
  \centering
  \begin{threeparttable}
  \tabcolsep3pt
  \caption{Characteristics of the Datasets.}
    \begin{tabular}{ccccccccc}
    \toprule[1.2pt]
    \multicolumn{1}{c}{{\bf WSC-2008's Datasets}}&{{\bf D-01}}&{{\bf D-02}}&{{\bf D-03}}&{{\bf D-04}}&{{\bf D-05}}&{{\bf D-06}}&{{\bf D-07}}&{{\bf D-08}}\cr
    \midrule[0.6pt]
    \multicolumn{1}{c}{\#Services}&158&558&604&1041&1090&2198&4113&8119\cr
    \midrule[0.6pt]
    \multicolumn{1}{c}{\#Concepts}&1540&1565&3089&3135&3067&12468&3075&12337\cr
    \midrule[0.6pt]
    \multicolumn{1}{c}{\#Sol.Services}&10&5&40&10&20&40&20&30\cr
    \bottomrule[1.2pt]
    \end{tabular}
    \end{threeparttable}
\end{table}

To validate our composition algorithm, we compared our approach with three different approaches in the same experimental environment. For each dataset, we mainly focused on the number of services in the composition result ({\it \#C.Services}) and the execution time to extract the solution from the service dependency graph ({\it C.Time}). The results are shown in Table 3.

\begin{table}[h]
  \centering
  \begin{threeparttable}
  \tabcolsep2.5pt
  \caption{Comparison with other approaches.}
  \label{tab:performance_comparison}
    \begin{tabular}{cccccccccc}
    \toprule[1.2pt]
    \multicolumn{2}{c}{{\bf Datasets}}&{{\bf D-01}}&{{\bf D-02}}&{{\bf D-03}}&{{\bf D-04}}&{{\bf D-05}}&{{\bf D-06}}&{{\bf D-07}}&{{\bf D-08}}\cr
    \midrule[0.6pt]
    \multirow{2}*{\bf Method in \cite{ref7}}&{\#C.Services}&10&5&40&10&20&35&20&30\cr
    {}&{C.Time (ms)}&47&78&1028&54&1295&137&243&191\cr
    \midrule[0.6pt]
    \multirow{2}*{\bf Method in \cite{ref10}}&{\#C.Services}&14&5&48&12&34&47&20&36\cr
    {}&{C.Time (ms)}&1&1&2&1&2&2&3&2\cr
    \midrule[0.6pt]
    \multirow{2}*{\bf Method in \cite{ref11}}&{\#C.Services}&10&5&40&10&20&35&20&30\cr
    {}&{C.Time (ms)}&61&52&176&122&156&855&193&304\cr
    \midrule[0.6pt]
    \multirow{2}*{\bf Our Method}&{\#C.Services}&\cellcolor[gray]{.8}\bf10&\cellcolor[gray]{.8}\bf5&\cellcolor[gray]{.8}\bf40&\cellcolor[gray]{.8}\bf10&\cellcolor[gray]{.8}\bf20&\cellcolor[gray]{.8}\bf35&\cellcolor[gray]{.8}\bf20&\cellcolor[gray]{.8}\bf30\cr
    {}&{C.Time (ms)}&6&10&21&13&22&61&33&20\cr
    \bottomrule[1.2pt]
    \end{tabular}
    \end{threeparttable}
\end{table}

As can be seen from Table 3, compared with other approaches, our approach can generate compositions with the same or smaller number of service. On the dataset {\it D-06}, our approach succeeds to find a better composition than the solution provided by the WSC'08 (35 versus 40). The execution time of \cite{ref10} is no more than 3 ms on all datasets, which proves that the method is efficient enough to solve the service composition problem. However, it always generates the compositions with more services than the others. Considering that all the methods except \cite{ref10} can find the minimal compositions, we compare our method with those methods in terms of the execution time.

\begin{figure}[b]
\centering
\includegraphics[width=4.6in]{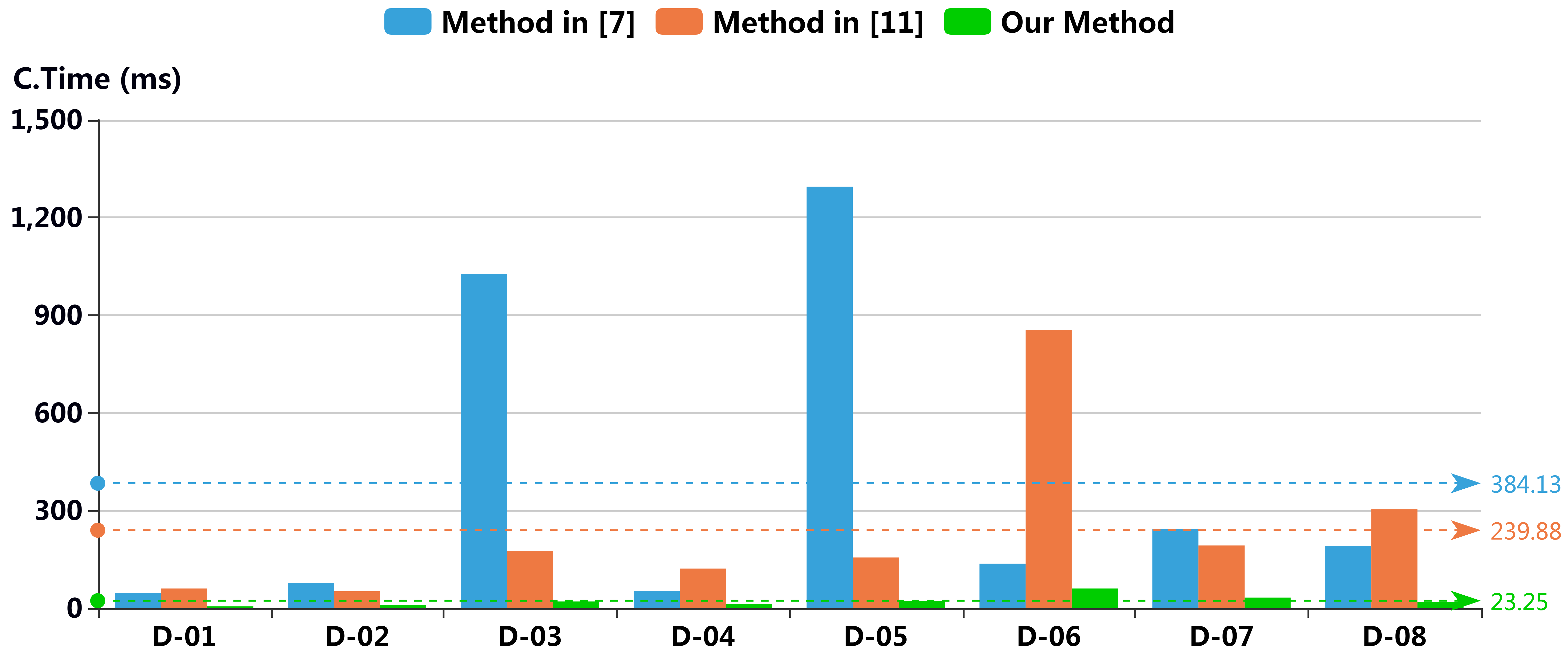}
\caption{Efficiency comparison.}
\end{figure}

Figure 10 shows that, our algorithm takes far less time to generate solutions than the other two. The dotted lines in blue and orange represents the average execution time of \cite{ref7} and \cite{ref11} respectively, while the green one shows the average time of our algorithm. Apparently our algorithm is over 10 times faster than \cite{ref7} and nearly 17 times faster than \cite{ref11} on average, which is a significant improvement.

\begin{table}[h]
  \centering
  \begin{threeparttable}
  \tabcolsep2.5pt
  \caption{Further comparison considering the service dependency graph.}
  \label{tab:performance_comparison}
    \begin{tabular}{cccccccccc}
    \toprule[1.2pt]
    \multicolumn{2}{c}{{\bf Datasets}}&{{\bf D-01}}&{{\bf D-02}}&{{\bf D-03}}&{{\bf D-04}}&{{\bf D-05}}&{{\bf D-06}}&{{\bf D-07}}&{{\bf D-08}}\cr
    \midrule[0.6pt]
    \multirow{3}*{\bf Method in \cite{ref7}}&{G.Size}&17&19&60&31&62&95&89&78\cr
    {}&{G.Time (ms)}&37&43&872&219&4861&536&7533&4761\cr
    {}&{Tot.Time (ms)}&84&121&1900&273&6156&673&7776&4952\cr
    \midrule[0.6pt]
    \multirow{3}*{\bf Method in \cite{ref11}}&{G.Size}&13&13&40&25&52&75&70&58\cr
    {}&{G.Time (ms)}&138&297&553&472&618&891&1253&1374\cr
    {}&{Tot.Time (ms)}&199&349&729&594&774&1746&1446&1678\cr
    \midrule[0.6pt]
    \multirow{3}*{\bf Our Method}&{G.Size}&60&61&104&43&101&170&140&124\cr
    {}&{G.Time (ms)}&5&12&62&15&62&181&403&576\cr
    {}&{Tot.Time (ms)}&11&22&83&28&84&242&436&596\cr
    \bottomrule[1.2pt]
    \end{tabular}
    \end{threeparttable}
\end{table}

We further compare our method with \cite{ref7} and \cite{ref11} taking the generation of the service dependency graph into account. As shown in Table 4, the size of the graph ({\it G.Size}) generated by \cite{ref7} and \cite{ref11} is smaller than ours because many optimizations are applied to reduce the graph size in these two methods, which leads to the fact that the time taken to generate the graph ({\it G.Time}) is longer than ours. Therefore, our knapsack-variant algorithm is executed over the the graph with lager size but is still over $10\times$ faster than the other two, which sufficiently indicates the efficiency of our composition algorithm. Even though taking the {\it G.Time} into consideration, the total time ({\it Tot.Time = G.Time + C.Time}) of our mechanism is still much less than the others. As a result, our mechanism is more applicable to the large-scale or real-time scenarios.

\section{Conclusions}

In this paper we proposed an effective and efficient mechanism to automatically generate the minimal composition over a service dependency graph. Each search step on the graph is ingeniously transformed into a dynamic knapsack problem, after which the proposed knapsack-variant algorithm is executed to minimize the number of services by solving each dynamic knapsack problem. Moreover, a full validation performed on eight different datasets from Web Service Challenge 2008 shows that our algorithm outperforms the state-of-the-art methods, as it executes much faster than the state-of-the-arts while keeping the minimal composition results and is applicable to the large-scale or real-time scenarios.

\section{Acknowledgment}

This work is funded by the Natural Science Foundation of China (Nos. 61673204, 61321491), State Grid Corporation of Science and Technology Projects (Funded No. SGLNXT00DKJS1700166), the Program for Distinguished Talents of Jiangsu Province, China (No. 2013-XXRJ-018), and the Fundamental Research Funds for the Central Universities (No. 020214380026).

\bibliographystyle{splncs}  
\bibliography{ref}

\end{document}